\documentclass[journal]{IEEEtran}
\usepackage{cite}
\usepackage{graphicx}
\usepackage{booktabs}

\usepackage{amsmath}
\usepackage{amssymb}
\usepackage{algorithm}
\usepackage{algorithmic}
\usepackage{url}
\begin{document}
\title{Bare Demo of IEEEtran.cls\\ for IEEE Journals}

\author{Dinesh Kumar Murugan,
        Nithyanandan Kanagaraj% <-this % stops a space
\thanks{\textbf{First Author:} Dinesh Kumar Murugan is from the Department of Physics, Indian Institute of Technology Hyderabad, Hyderabad, India – 502285.}% <-this % stops a space
\thanks{\textbf{Corresponding author:} Nithyanandan Kanagaraj is an Assistant Professor and leads the Ultrafast Fiber Optics \& Smart Photonic Technologies Lab, Department of Physics, Indian Institute of Technology Hyderabad, India – 502285. (email: nithyan@phy.iith.ac.in)}}


\title{Bidirectional Fourier-Enhanced Deep Operator Network for Spatio-Temporal Propagation in Multi-Mode Fibers}



\maketitle

\begin{abstract}
	Ultrashort-pulse propagation in graded-index multimode fibers is a highly nonlinear phenomenon driven by several physical processes. Although conventional numerical solvers can reproduce this behavior with high fidelity, their computational cost limits real-time prediction, rapid parameter exploration, experimental feedback, and especially inverse retrieval of input fields from measured outputs. In this work, we introduce an operator learning framework that learns both the forward and inverse propagation operators within a single unified architecture. By combining spectral filters for spatio-temporal representations with Fourier-embedded conditioning on physical parameters, the model functions as a fast surrogate capable of accurately predicting complex field evolution on previously unseen cases. To our knowledge, this represents one of the first demonstrations of a bidirectional operator-learning framework applied to ultrashort-pulse multimode fiber propagation. The resulting architecture enables orders-of-magnitude speedup over numerical solvers, paving the way for real-time beam diagnostics, data-driven design of complex input fields, and closed-loop spatio-temporal control. Moreover, the same framework can potentially be applied to a wide variety of wave systems exhibiting analogous nonlinear and dispersive effects in optics and beyond.
\end{abstract}


% \fi
\IEEEpeerreviewmaketitle

\section{Introduction}
The propagation of intense ultrashort pulses in \textbf{Gr}aded-\textbf{In}dex (GRIN) \textbf{M}ulti-\textbf{M}ode \textbf{F}ibers (MMFs) gives rise to some of the most complex spatio-temporal nonlinear phenomena in optics. Kerr-driven self-focusing, geometric parametric instability, massive intermodal coupling, multimode dispersion, and spatio-temporal instabilities combine to produce rapidly evolving transverse speckle patterns and temporal pulse reshaping over millimeter scale distances~\cite{agrawal2013nonlinear,AGRAWAL2019309,Wright2015,krupa2016observation,8094262,krupa2017spatiotemporal,10.1063/1.5119434,Hansson:20}. These dynamics underpin MMF-based high-power beam delivery, endoscopic imaging, spatio-temporal mode-locked lasers, and emerging neuromorphic optical computing platforms~\cite{Krupa_2017,Ai:17,Morales-Delgado:15,Gusachenko2017,doi:10.1126/science.aao0831,PhysRevLett.126.093901,Cao2023,Paudel_2020,Rahmani:21}. Yet full (3+1)D (3 spatial coordinates and 1 time coordinate) modeling using methods like \textbf{S}plit-\textbf{S}tep \textbf{F}ourier \textbf{M}ethod (SSFM) or \textbf{F}inite \textbf{D}ifference - \textbf{B}eam \textbf{P}ropagation \textbf{M}ethod (FD-BPM) is computationally expensive: needing GPU or cluster acceleration, and offer no native capability for inverse design \cite{kolesik2017computational,Brehler_2020}.\smallskip

In recent years, \textbf{D}eep \textbf{L}earning (DL) has emerged as a powerful alternative for accelerating the solution of complex nonlinear optical systems~\cite{8437734,Genty2021,9596142,Salmela2021,teugin2021reusability,Chen2023,Fang:23,Hadad_2023,10.1063/5.0169810,belonovskii2024vcsel,10.1002/adpr.202500149,Freire:23}. \textbf{P}hysics-\textbf{I}nformed \textbf{N}eural \textbf{N}etworks (PINNs), \textbf{F}ourier \textbf{N}eural \textbf{O}perators (FNOs), and \textbf{Deep} \textbf{O}perator \textbf{Net}works (DeepONets) have all demonstrated impressive performance in simulating physical systems~\cite{RAISSI2019686,Lu2021,li2021fourierneuraloperatorparametric,10.5555/3495724.3496356}. However, operator-learning frameworks, like FNOs, and DeepONets, remain largely unidirectional, and have only very recently begun to be applied to nonlinear optical systems~\cite{MARGENBERG2024112725,long2025invertiblefourierneuraloperators,9815178,10195160}.\smallskip

Practical applications in multimode systems increasingly demand not only rapid forward prediction but also robust inverse retrieval of launch conditions from downstream measurements. Conventional numerical schemes lack a well-defined inverse operator and rely on slow iterative optimization or heuristic search, which become unstable in the strongly nonlinear regime~\cite{10.1364/AOP.484298,10.1117/1.APN.2.6.066005}. Operator-learning frameworks, by mapping directly between function spaces rather than finite parameter vectors, provide a natural paradigm for representing field evolution governed by nonlinear physics. FNOs along with DeepONets, are well suited to model systems mainly shaped by dispersive and diffractive effects.\smallskip

Motivated by these challenges, we present a bidirectional Fourier-enhanced DeepONet that learns both the forward propagation operator and its inverse within a single unified architecture. By jointly training on paired input-output intensity observations and conditioning on propagation distance and peak power, the proposed model serves as a flexible surrogate for the underlying nonlinear dynamics. This framework enables microsecond-scale forward and inverse operation, opening new possibilities for real-time experimental feedback, adaptive beam shaping, and data-driven design in multimode fibre systems while remaining readily extendable to other wave-propagation problems exhibiting analogous nonlinear and dispersive interactions.\smallskip

The following work is further organized into six sections. \textbf{Section \ref{simulation-dataset}: Nonlinear Propagation Simulation and Data Generation} outlines the GRIN MMF dataset generation pipeline, including GNLSE-based simulations, bidirectional pairing, and preparing it for training. \textbf{Section \ref{operator-learning}: Modeling Nonlinear Propagation using Operator Learning Networks} outlines the design, architecture, and training strategy of the bidirectional Fourier-enhanced DeepONet. \textbf{Section \ref{results}: Results} presents the results demonstrating the model’s accuracy in forward and inverse multimode propagation. \textbf{Section \ref{discussion}: Discussion} discusses the physical relevance of the learned operator and outlines directions for future improvement. \textbf{Section \ref{conclusion}: Conclusion} summarizes the main findings and highlights the potential of the proposed operator-learning framework. \textbf{Appendix \ref{app:pseudocode}: Pseudo-code for the Architecture} presents the algorithmic structure of the proposed model, while \textbf{Appendix \ref{app:methods}: Experimental Methods} describes the simulation methods and neural-network building blocks used in the study.

\section{Nonlinear Propagation Simulation and Data Generation} \label{simulation-dataset}

The dataset used in this work is generated using a (3+1)D SSFM solver for the \textbf{G}eneralized \textbf{N}on\textbf{L}inear \textbf{S}chrödinger \textbf{E}quation (GNLSE) in GRIN MMFs. The numerical implementation is adapted from the open-source code accompanying the work of Teğin et al \cite{teugin2021reusability}. We gratefully acknowledge their contribution and the public availability of their simulation framework, which has significantly facilitated reproducible research in spatio-temporal multimode fibre optics.

\subsection{Governing Equation}

Like mentioned previously, the propagation of ultrashort optical pulses in a graded-index multimode fiber is governed by the GNLSE. For a multimode field \(A(x,y,t,z)\), the equation reads

\begin{equation} \label{eq:gnlse}
	\frac{\partial A}{\partial z} = \hat{D} A + i \gamma |A|^2 A,
\end{equation}
where \(z\) is the propagation distance, \(\gamma = 2 \pi n_2 / \lambda_c\) is the Kerr nonlinear coefficient with \(n_2 = 3.2\times 10^{-20}\,\mathrm{m^2/W}\), and \(\lambda_c = 1030\,\mathrm{nm}\) is the central wavelength. The linear operator \(\hat{D}\) accounts for diffraction and chromatic dispersion:

\begin{equation}
	\hat{D} = \frac{i}{2 k_0} \nabla_\perp^2 + i \frac{\beta_2}{2} \frac{\partial^2}{\partial t^2} + i \frac{\beta_3}{6} \frac{\partial^3}{\partial t^3},
\end{equation}
where \(\nabla_\perp^2 = \partial^2/\partial x^2 + \partial^2/\partial y^2\) is the transverse Laplacian, \(k_0 = 2 \pi n_0 / \lambda_c\) is the propagation constant for the fiber core with refractive index \(n_0 = 1.45\), \(\beta_2 = 24.8\,\mathrm{fs^2/mm}\) and \(\beta_3 = 23.3\,\mathrm{fs^3/mm}\) are the second- and third-order dispersion coefficients. The fiber radius is \(R = 25\,\mu\mathrm{m}\) and the relative index difference is \(\Delta = 0.01\).\smallskip

The nonlinear term \(i \gamma |A|^2 A\) represents the instantaneous Kerr effect, and the parabolic index profile leads to self-imaging with a period

\[
s_{\mathrm{img}} = \frac{\pi R}{\sqrt{2 \Delta}} \approx 5.55\times10^{-4}\,\mathrm{m}.
\]

The transverse window spans \(54.144\,\mu\mathrm{m} \times 54.144\,\mu\mathrm{m}\) on a \(64\times64\) grid, and the temporal window has width \(1.8\,\mathrm{ps}\) sampled with 1024 points. These resolutions ensure that nonlinear spatial reshaping, mode beating, and temporal dynamics are accurately resolved~\cite{teugin2021reusability,Mafi:10,snyder1983optical,agrawal2010applications}.\smallskip 

The input pulse consists of a 100 fs Gaussian envelope multiplied by a normalized random superposition of the six lowest-order fiber modes. The peak power \(P_0\) is swept from \(1\,\mathrm{MW}\) to \(1\,\mathrm{GW}\), uniformly sampled over 300 values, in order to span weakly to strongly nonlinear propagation regimes. The total fiber length is set to \(10\,s_{\mathrm{img}} \approx 5.55\,\mathrm{mm}\), and the propagation is discretized into 480 steps along \(z\). At each propagation step, the transverse and temporal intensities are recorded:

\[
I(x,y;z) = \int |A(x,y,t,z)|^2 \, dt
\]

\[
I(t;z) = \iint |A(x,y,t,z)|^2 \, dx\, dy
\]
which form the basis for training the bidirectional operator-learning model.

\subsection{Bidirectional Dataset Construction}\label{dataset}
To enable a single neural operator to learn both the propagation
\[
G\!:\; \{I(x,y;0), I(t;0), P_{0}, z\} \rightarrow \{I(x,y;z), I(t;z)\}
\]
and the corresponding inverse operation,
\[
G^{-1}\!:\; \{I(x,y;z), I(t;z), P_{0}, z\} \rightarrow \{I(x,y;0), I(t;0)\}
\]
each simulated propagation trajectory is augmented with its reversed counterpart. In this reversed pair, the field observed at propagation distance \(z\) becomes the input, while the initial field at \(z=0\) becomes the target. The peak power \(P_{0}\) and propagation distance \(z\) are retained as conditioning variables. A binary direction indicator \(d \in \{0,1\}\) is concatenated to the input, where \(d = 0\) denotes forward operation and \(d = 1\) denotes inverse operation.\smallskip

This augmentation doubles the effective dataset size and allows the neural operator to perform both forward propagation and inverse operation of the full-field \((3+1)\)D spatio-temporal evolution.

\subsection{Dataset Preparation}
After bidirectional augmentation, the dataset is prepared for neural network training using two main steps:

\begin{itemize}
	\item \textbf{Dataset Splitting:}  
	The augmented dataset is divided into training and validation subsets using a random stratified 80\%–20\% split, ensuring balanced representation during optimization. A separate test set is generated independently to make sure that final evaluation is performed on completely unseen data.
	
	\item \textbf{Scaling:}  
	All intensity fields, peak powers, and propagation distances are scaled using standard normalization based on the training set only to avoid data leakage. Each feature $x$ is transformed according to
	\begin{equation}
		x' = \frac{x - \mu}{\sigma + \epsilon}
	\end{equation}
	where $\mu$ and $\sigma$ denote the mean and standard deviation of the feature computed over the training set, and $\epsilon = 10^{-8}$ prevents division by zero. This normalization improves numerical stability and ensures that all input parameters contribute on comparable scales during model training.
\end{itemize}

These preparation steps enhance the efficiency and stability of training, and improve the model's ability to generalize to unseen conditions~\cite{MAHARANA202291}.

\begin{figure*}[ht!]
	\centering
	\includegraphics[width=0.95\textwidth]{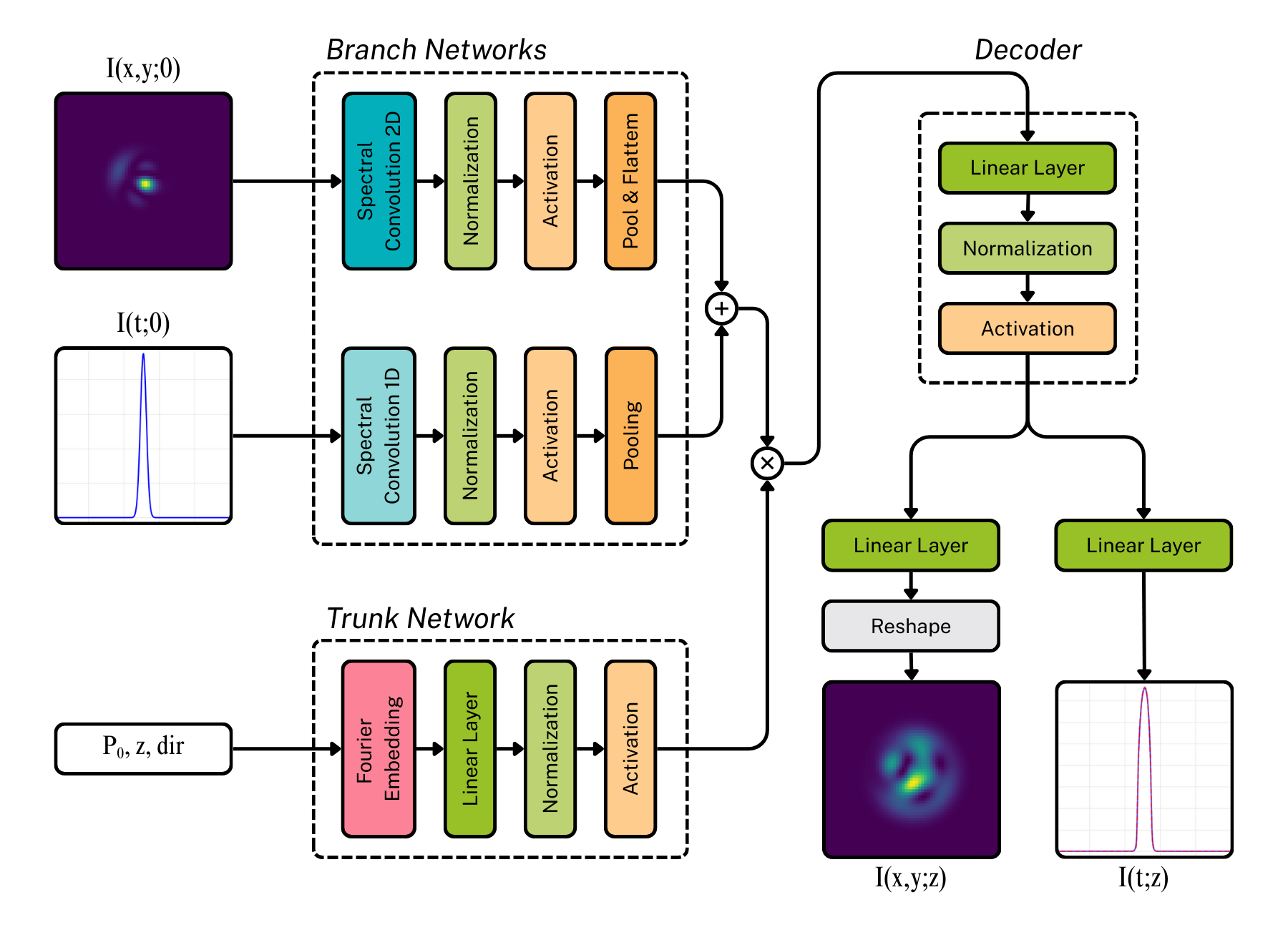}
	\caption{Schematic of the proposed bidirectional Fourier-enhanced DeepONet for joint forward and inverse modeling of nonlinear pulse propagation in graded-index multimode fibers. 
		Two dedicated branch networks separately process the input transverse intensity distribution $I(x,y;0)$ (or $I(x,y;z)$) and temporal intensity trace $I(t;0)$ (or $I(t;z)$) using spectral convolutions in their respective domains. 
		The resulting latent representations are concatenated (denoted by $\oplus$) and combined via element-wise multiplication (denoted by $\otimes$) with the output of a trunk network that encodes the physical parameters (peak power $P_0$, propagation distance $z$, and a binary direction flag $d \in \{0,1\}$ indicating forward or inverse operation) through Fourier feature embeddings. 
		A shared decoder followed by two output projections generates the predicted transverse speckle pattern $I(x,y;z)$ and temporal trace $I(t;z)$ (forward mode) or recovers the initial state $I(x,y;0)$ and $I(t;0)$ (inverse mode). 
		The depth, width, and number of retained Fourier modes in each branch were adjusted empirically to balance expressive power and computational efficiency for the present (3+1)D multimode propagation task.}
	\label{fig:architecture}
\end{figure*}

\section{Modeling Nonlinear Propagation using Operator Learning Networks} \label{operator-learning}

\begin{figure*}[ht!]
	\centering
	\includegraphics[width=\textwidth]{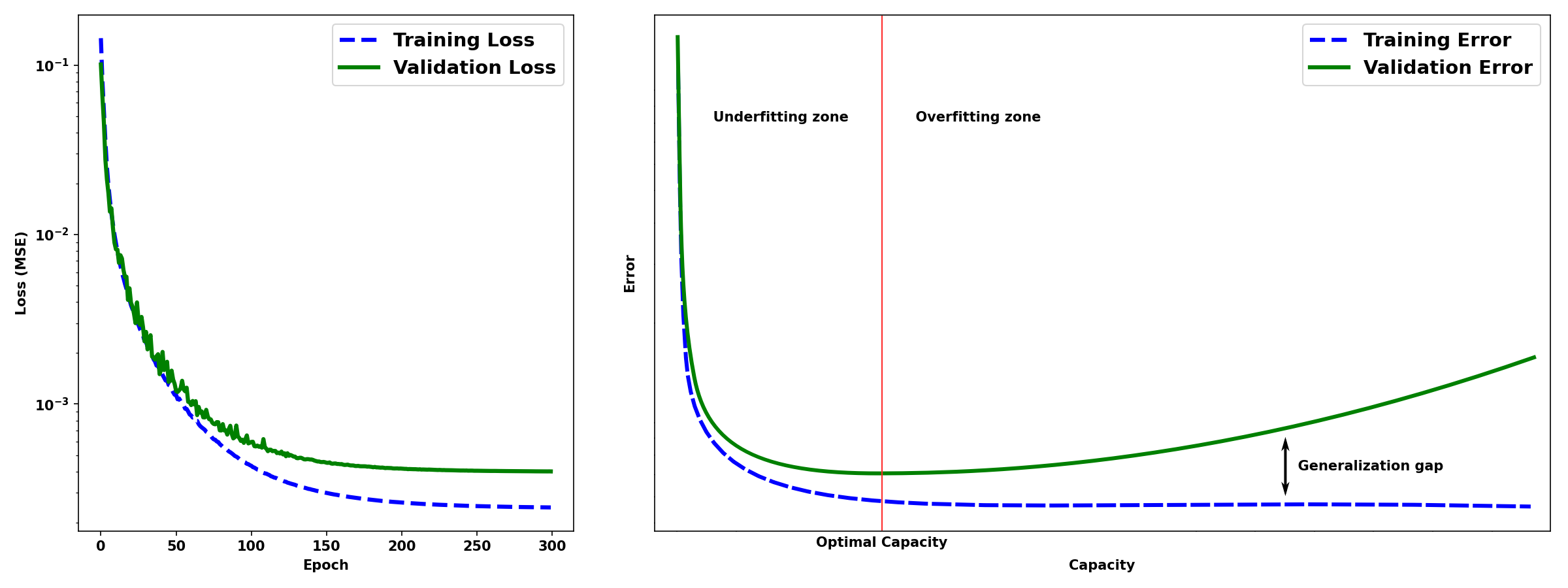}
	\caption{\textit{Left:} Training and validation loss curves of the model, showing stable convergence and a small generalization gap at the selected checkpoint. \textit{Right:} Illustration adapted from Goodfellow et al.~\cite{Goodfellow-et-al-2016} showing the relationship between model capacity and generalization error, indicating the ideal stopping point prior to overfitting. The behavior observed in the left panel indicates that the model is stopped around this optimal zone.}
	\label{fig:training_curve}
\end{figure*}

\subsection{Architecture} \label{acrhitecture}
The standard DeepONet approximates nonlinear operators by combining a branch network, which processes discrete samples of the input function $u(x)$, and a trunk network, which processes the evaluation coordinates $s$. The output is expressed as
\begin{equation}
	G(u)(s) = \sum_{k=1}^{p} b_k(u) \, t_k(s)
\end{equation}
where $b_k(u)$ are the features learned by the branch network and $t_k(s)$ are the outputs of the trunk network forming the basis functions~\cite{Lu2021}.\smallskip

To better capture global correlations and oscillatory features, we enhance the standard DeepONet using Fourier-based modifications:

\begin{enumerate}
	\item \textbf{Fourier Branch Network:}  
	The branch network is replaced with a spectral convolution module inspired by FNOs. The discrete input function $u(x)$ is transformed into Fourier space, filtered through learnable spectral weights, and then transformed back to produce the branch outputs $b_k^{\text{Fourier}}(u)$. This approach allows the network to capture long-range spatial correlations efficiently~\cite{li2021fourierneuraloperatorparametric}.
	
	\item \textbf{Fourier Trunk Network:}  
	The trunk network is augmented with a learned Fourier feature embedding. The physical parameters \(s\) (here representing power, distance, and direction) are projected through a bank of sinusoidal features using a learnable matrix, producing a high-dimensional periodic encoding \(t_k^{\text{Fourier}}(s)\). This enhances the expressiveness of the evaluation basis and enables the model to better represent oscillatory dependencies and nonlinear parameter interactions~\cite{10.5555/3495724.3496356}.
\end{enumerate}

With these enhancements, the output operator can be expressed in a form analogous to the standard DeepONet:
\begin{equation}
	G(u)(s) = \sum_{k=1}^{p} b_k^{\text{Fourier}}(u) \, t_k^{\text{Fourier}}(s)
\end{equation}
where the branch outputs are learned via spectral/Fourier layers and the trunk outputs correspond to Fourier features.\smallskip

The bidirectional Fourier-enhanced DeepONet implemented in this work combines 2D spectral convolutions for transverse speckles, 1D spectral convolutions for temporal traces, and a Fourier-embedded trunk network to process the physical parameters (peak power, propagation distance, and direction flag). The concatenated branch and trunk features are passed through a shared decoder, which splits into two separate output heads for spatial and temporal predictions. Figure~\ref{fig:architecture} provides a high-level schematic of the model, while a more detailed algorithmic description is given in Appendix~\ref{app:pseudocode}, enabling simultaneous learning of forward and inverse mappings within a single network.

\subsection{Training and Optimization}
\label{sec:training}
The training procedure is designed to make use of the symmetry of the dataset while maintaining stable convergence in a high-dimensional nonlinear regression problem. By constructing matched forward and inverse pairs, as discussed in Sub-Section~\ref{dataset}, the network sees both directions of the propagation operator within each batch. This removes the need for two separate models and allows a single set of parameters to learn a consistent mapping for both tasks. A binary direction flag is included as an additional input so the network can distinguish between forward prediction and inverse reconstruction without adding extra architectural complexity.\smallskip

The network is trained using the \textbf{Ada}ptive \textbf{m}omentum estimation with decoupled \textbf{W}eight-decay (AdamW) optimizer, which decouples weight decay from the gradient update and improves stability for deep regression models. Training begins with a learning rate of $5 \times 10^{-4}$, and an exponential decay factor of $\gamma=0.98$ is applied after each epoch to support gradual refinement. The model is trained for 300 epochs with a batch size of 64, and gradient clipping with a maximum $L_2$ norm of 1.0 is used to prevent occasional spikes caused by highly nonlinear samples.\smallskip

A joint mean-squared-error loss is used to optimize both spatial and temporal outputs simultaneously:
\begin{equation}
	\mathcal{L} \;=\; 
	\big\|\hat{I}_{s}^{\,\text{pred}} - \hat{I}_{s}^{\,\text{true}}\big\|_2^2
	\;+\;
	\big\|\hat{I}_{t}^{\,\text{pred}} - \hat{I}_{t}^{\,\text{true}}\big\|_2^2 
\end{equation}
where $\hat{I}_{s}$ and $\hat{I}_{t}$ denote the normalized spatial and temporal representations, respectively. This joint objective encourages the network to learn shared structure between the two representations and improves overall stability during training.\smallskip

Validation loss is recorded after each epoch, and the model checkpoint with the lowest value is used for final evaluation. This ensures that the selected network achieves the best generalization performance. The training curves, as seen in Figure ~\ref{fig:training_curve}, show a smooth decrease in both training and validation loss, indicating stable convergence without signs of overfitting~\cite{Goodfellow-et-al-2016,bishop2023deep}.

\section{Results} \label{results}

\begin{figure*}[ht!]
	\centering
	\includegraphics[width=\textwidth]{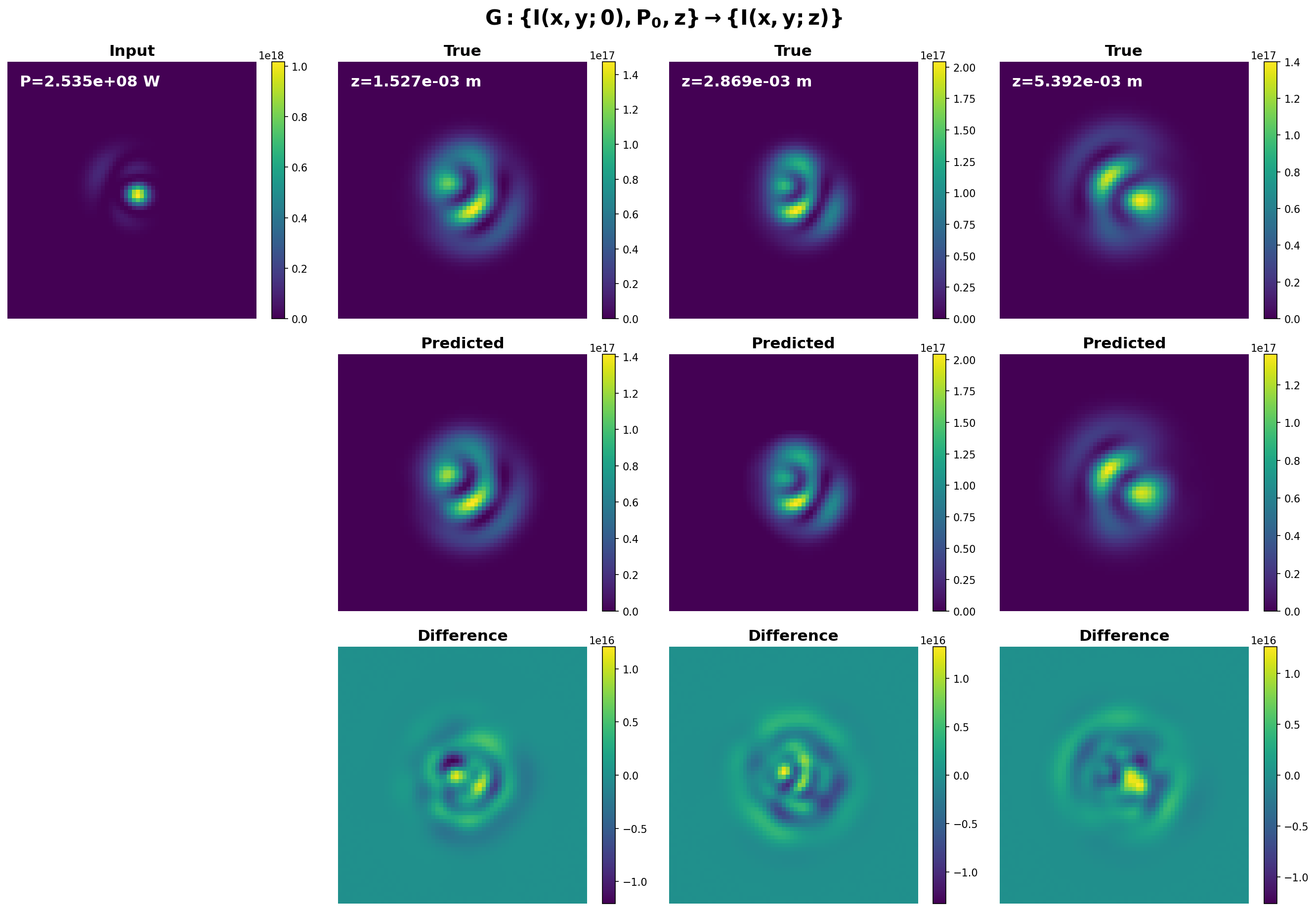}
	\caption{
		Forward propagation of transverse intensity $I(x,y;z)$ predicted by the trained network on a representative test sample (peak power $P_0 = 253.5$~MW) from the held-out test set.
		\textit{Leftmost plot:} input transverse intensity $I(x,y;0)$. 
		\textit{Top row plots:} ground-truth intensity distributions $I(x,y;z)$ at three propagation distances ($z = 1.527$~mm, $2.869$~mm, and $5.392$~mm). 
		\textit{Middle row plots:} corresponding predictions from the trained network. 
		\textit{Bottom row plots:} signed difference between true and predicted speckle patterns.
	}
	\label{fig:forward_speckle}
\end{figure*}

\begin{figure*}[ht!]
	\centering
	\includegraphics[width=\textwidth]{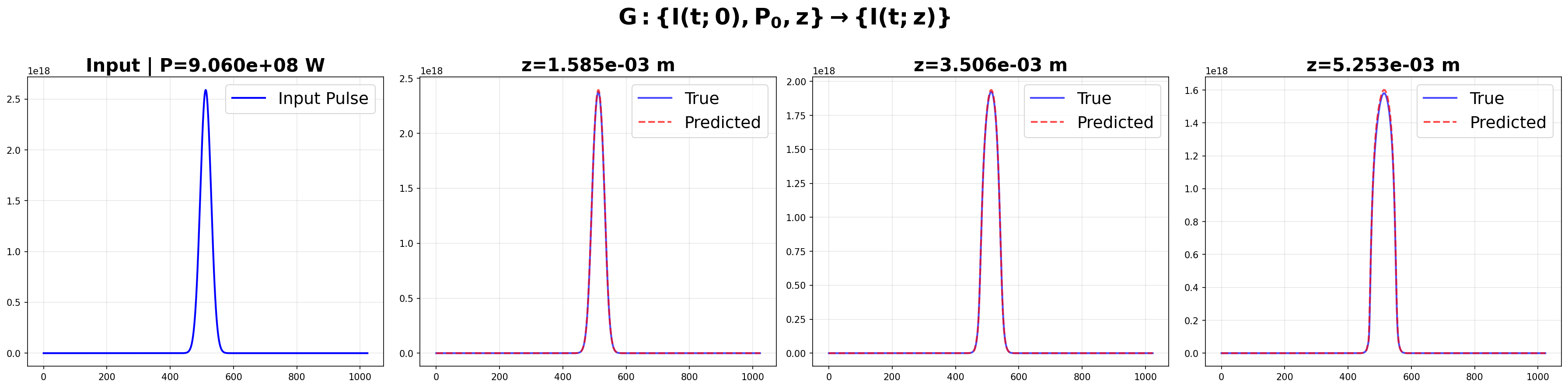}
	\caption{
		Forward propagation of the spatially integrated temporal intensity trace $I(t;z)$ for a representative test sample (peak power $P_0 = 906.0$~MW) from the held-out test set. 
		\textit{First plot} shows input trace $I(t;0)$. 
		\textit{The three right plots} show ground-truth traces (blue solid) at propagation distances $z = 1.585$~mm, $3.506$~mm, and $5.253$~mm, with network predictions (red dashed) overlaid. 
	}
	\label{fig:forward_trace}
\end{figure*}

\subsection{Forward Propagation} \label{forward}
The results show that the model predicts forward propagation accurately for both spatial and temporal intensities. As reported in Table~\ref{tab:forward_results}, the RMSE and MAE values are low for \(I(x,y;z)\) and \(I(t;z)\), indicating good agreement with the ground-truth data on the test set. This suggests that the network generalizes well to unseen propagation conditions.

\begin{table}[!ht]
\caption{Forward Propagation Performance on the Held-Out Test Set (Normalized Intensity Scale).}
\label{tab:forward_results}
\centering
\begin{tabular}{lcc}
\toprule
& \textbf{RMSE} & \textbf{MAE}\\
\midrule
$I(x,y;z)$ & $9.372\times10^{-2}$ & $1.954\times10^{-2}$\\
$I(t;z)$ & $1.359\times10^{-2}$ & $2.200\times10^{-3}$\\
\bottomrule
\end{tabular}
\end{table}

Visual comparisons further support this observation. As shown in Fig.~\ref{fig:forward_speckle} and Fig.~\ref{fig:forward_trace}, the predicted spatial speckle patterns and temporal intensity traces very closely follow the corresponding reference simulations. Both the overall structure and fine details are well preserved, confirming the quality of the forward predictions.

\begin{figure*}[!ht]
	\centering
	\includegraphics[width=\textwidth]{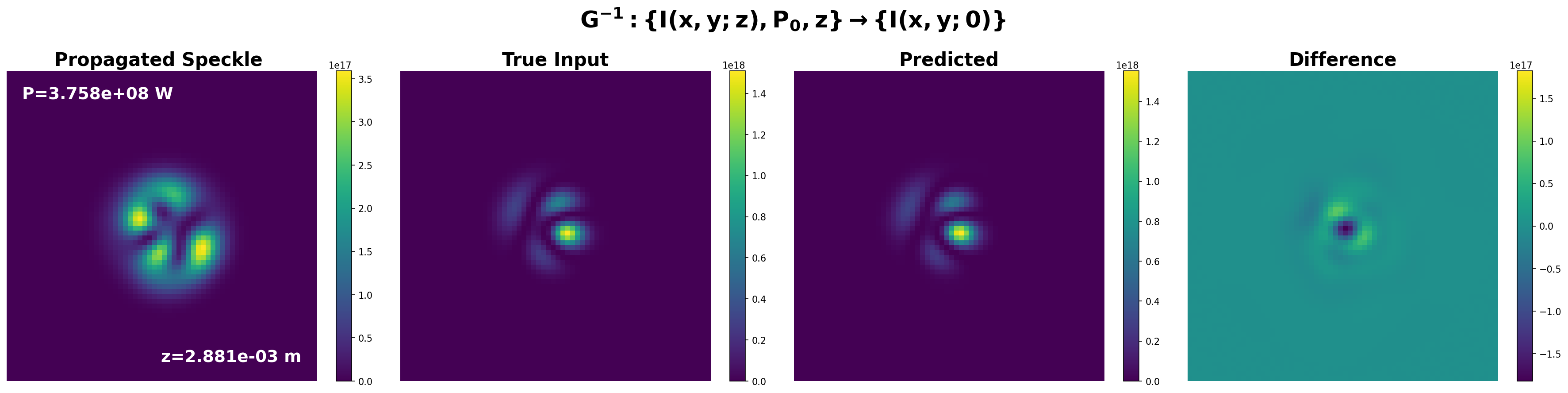}
	\caption{
		Inverse operation of the input transverse intensity $I(x,y;0)$ from a propagated observation in the held-out test set (peak power $P_0 = 375.8$~MW, observation distance $z = 2.881$~mm). 
		\textit{First plot:} propagated speckle pattern $I(x,y;z)$ used as input to the network. 
		\textit{Second plot:} ground-truth initial intensity $I(x,y;0)$. 
		\textit{Third plot:} network prediction conditioned only on the observed speckle, peak power $P_0$, and distance $z$ (inverse mode, direction flag $d=1$). 
		\textit{Fourth plot:} signed difference (prediction minus ground truth). 
	}
	\label{fig:inverse_speckle}
\end{figure*}

\begin{figure*}[!ht]
	\centering
	\includegraphics[width=\textwidth]{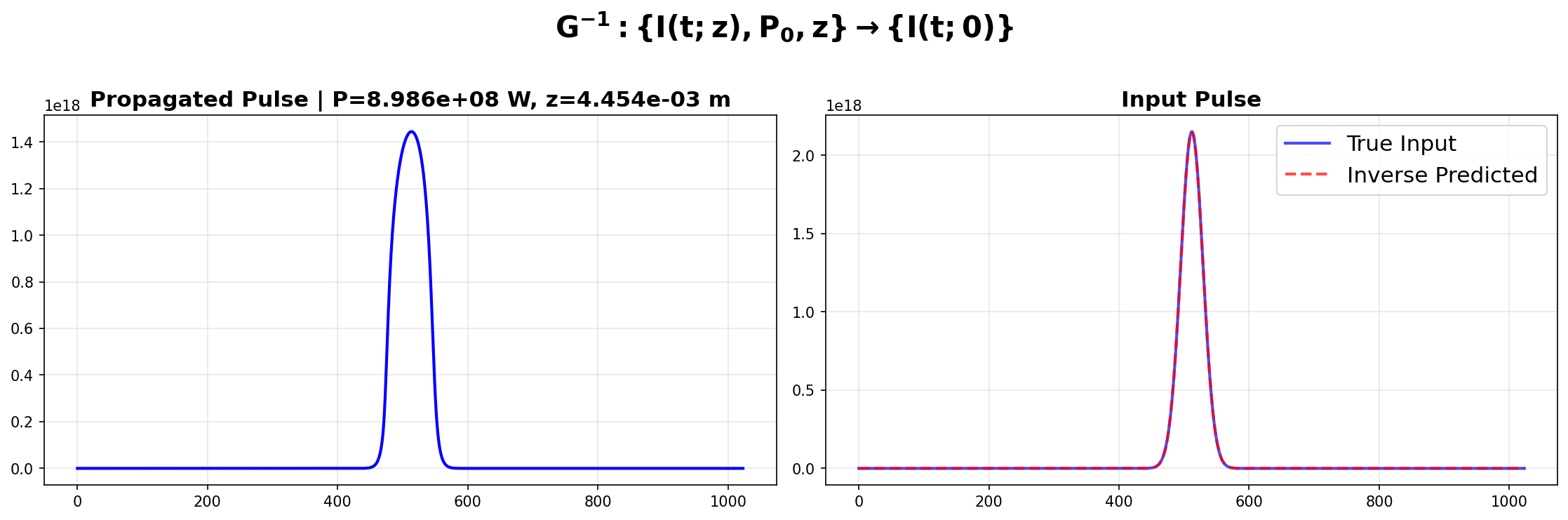}
	\caption{
		Inverse reconstruction of the initial temporal intensity trace $I(t;0)$ from a propagated observation in the held-out test set (peak power $P_0 = 898.6$~MW, observation distance $z = 4.454$~mm). 
		\textit{First plot:} propagated temporal trace $I(t;z)$ used as input to the network. 
		\textit{Second plot:} ground-truth input trace $I(t;0)$ (blue solid) with the network’s inverse prediction (red dashed) overlaid. 
	}
	\label{fig:inverse_trace}
\end{figure*}

\subsection{Inverse Operation} \label{inverse}
The inverse results demonstrate that the model is also able to recover the input fields with good accuracy. As summarized in Table~\ref{tab:inverse_results}, the error values remain low for both the reconstructed spatial intensity \(I(x,y;0)\) and the temporal intensity \(I(t;0)\) on the held-out test set. While the inverse task is inherently more challenging than forward prediction, the obtained metrics indicate reliable reconstruction performance.

\begin{table}[!ht]
\caption{Inverse Operation Performance on the Held-Out Test Set (Normalized Intensity Scale).}
\label{tab:inverse_results}
\centering
\begin{tabular}{lcc}
\toprule
& \textbf{RMSE} & \textbf{MAE}\\
\midrule
$I(x,y;0)$ & $1.872\times10^{-1}$ & $3.155\times10^{-2}$\\
$I(t;0)$ & $1.366\times10^{-2}$ & $1.943\times10^{-3}$\\
\bottomrule
\end{tabular}
\end{table}

This behavior is further confirmed by visual inspection. Figures~\ref{fig:inverse_speckle} and \ref{fig:inverse_trace} show that the reconstructed spatial speckle patterns and temporal intensity traces closely match the reference inputs. The main structural features are well captured, demonstrating the network’s ability to perform stable and meaningful inverse operation.

\section{Discussion} \label{discussion} 
An important aspect of this work is how the learned operator begins to reflect the underlying physics of nonlinear multimode propagation. Although the network is trained directly on intensity data without explicit access to phase or modal amplitudes, it still internalizes the characteristic signatures of Kerr-induced spatio-temporal coupling, modal interference, and longitudinal evolution. The consistency of its predictions across a wide range of input powers suggests that the model captures stable structural relationships that are normally described by the GNLSE. This offers a promising direction for using operator-learning networks as data-driven surrogates for complex nonlinear propagation models while preserving essential physical behavior.\smallskip

The implications of this extend beyond fast simulation. The ability to recover the initial field from a single propagated measurement shows that the network learns an approximate inverse of the nonlinear operator, which is analytically intractable for multimode fibers. This suggests potential applications in multimode pulse characterization, nonlinear imaging, and digital correction of modal distortions. The fact that both the forward and inverse mappings are handled by a single coherent operator model is especially notable, since it shows that the network is not simply fitting pointwise relationships, but rather learning a structured representation of the entire propagation process.\smallskip

A natural next step is to move from simulation-only training to learning directly from experimental data. Real measurements introduce noise, modal coupling imperfections, and alignment variability that are not fully captured in numerical propagation. Incorporating physics-based regularization terms inspired by PINN frameworks could help constrain the model when experimental supervision is limited. This combination of real data and physically guided loss terms would likely improve the reliability of both forward predictions and inverse reconstructions.\smallskip

Complementary to this, improving robustness to incomplete or noisy measurements would further support experimental deployment. In practice, spatial and temporal detectors often differ in resolution and noise characteristics, and some modalities may be unavailable in certain settings. Jointly training on multiple measurement types, such as speckle images together with temporal traces or phase-sensitive signals, could help the network better separate modal interactions and reduce ambiguity in the inverse problem.\smallskip

The current work also focuses on a specific fiber geometry and propagation regime. Extending the training distribution to include different refractive index profiles, modal sets, nonlinear coefficients, and pulse bandwidths would allow the model to operate more flexibly across a broader class of multimode fibers. Techniques such as domain randomization or meta-learning could enable the network to adapt quickly to new fibers using only a small set of calibration examples.\smallskip

Finally, future work should focus on extending this framework toward higher spatial and temporal resolution. Although the propagation distance is modeled continuously, the spatial and temporal dimensions are still discretized. Further development of the operator-learning formulation, using implicit or continuous representations in space and time, could enable effectively infinite-resolution predictions.

\section{Conclusion} \label{conclusion} 

In this work, we demonstrated that the proposed operator-learning framework can accurately and efficiently model nonlinear spatio-temporal pulse propagation in GRIN MMFs. Compared to conventional numerical solvers, the learned network provides a substantial computational advantage. While a single forward propagation using the SSFM requires \(8.646\,\mathrm{ms}\), the trained network performs forward prediction in \(9.650\times10^{-2}\,\mathrm{ms}\) and inverse reconstruction in \(9.249\times10^{-2}\,\mathrm{ms}\), offering an almost 90 times speedup. All timings were obtained on the same NVIDIA P100 GPU, ensuring a fair comparison.\smallskip

Beyond speed, the model achieves low prediction errors across both spatial and temporal intensity representations. As shown by the quantitative metrics and visual comparisons, the network faithfully reproduces the key features of the propagated fields in the forward direction and remains stable when solving the inverse problem, which is traditionally ill-posed and computationally demanding. This level of accuracy indicates that the learned operator captures the essential physics of nonlinear multimode propagation rather than merely interpolating the training data.\smallskip

To reiterate our discussion, the combination of high accuracy, strong generalization, and orders-of-magnitude acceleration makes this approach attractive for a wide range of real-world applications. While demonstrated here for nonlinear multimode fiber propagation, the unified forward and inverse operator-learning framework is not restricted to optical systems and can be extended to other physical processes governed by complex partial differential equations. The ability to perform fast simulation and inverse reconstruction within a single model enables real-time system design, rapid parameter exploration, and experimental feedback or control in settings where conventional numerical solvers are prohibitively slow.

\appendices
\section{Pseudo-code for the Architecture}\label{app:pseudocode}
\begin{algorithm}[!ht]
\caption{Bidirectional Fourier-Enhanced DeepONet}
\begin{algorithmic}[1]
\STATE \textbf{Input:} Spatial field $\mathbf{u}_s \in \mathbb{R}^{H \times W}$, Temporal field $\mathbf{u}_t \in \mathbb{R}^{T}$, Physical parameters $\boldsymbol{\theta} = [P_0, z, d]$ where $d=0$ for forward and $d=1$ for inverse operation
\STATE \textbf{Hyperparameters:} Latent dimension $p$, Fourier modes $(m_{2d}, m_{1d})$, Hidden dimension $d_h$, Number of spectral blocks $N_s, N_t$\smallskip
\STATE \textbf{1. Spatial Branch Network} $\mathcal{B}_s$:
\STATE \hspace{1em} Initialize $\mathbf{z}_s = \mathbf{u}_s$
\STATE \hspace{1em} For $i = 1$ to $N_s$:
\STATE \hspace{2em} Apply SpectralConv2D($m_{2d}$) $\rightarrow$ LayerNorm $\rightarrow$ Activation $\rightarrow$ MaxPool2D
\STATE \hspace{1em} Flatten and project through MLP to latent vector $\mathbf{b}_s \in \mathbb{R}^p$\smallskip
\STATE \textbf{2. Temporal Branch Network} $\mathcal{B}_t$:
\STATE \hspace{1em} Initialize $\mathbf{z}_t = \mathbf{u}_t$
\STATE \hspace{1em} For $i = 1$ to $N_t$:
\STATE \hspace{2em} Apply SpectralConv1D($m_{1d}$) $\rightarrow$ LayerNorm $\rightarrow$ Activation $\rightarrow$ MaxPool1D
\STATE \hspace{1em} Flatten and project through MLP to latent vector $\mathbf{b}_t \in \mathbb{R}^p$\smallskip
\STATE \textbf{3. Trunk Network} $\mathcal{T}$:
\STATE \hspace{1em} Apply Fourier feature embedding to $\boldsymbol{\theta}$, then MLP to $\mathbf{t} \in \mathbb{R}^{2p}$\smallskip
\STATE \textbf{4. Combine Branches and Trunk:} 
\STATE \hspace{1em} Concatenate branch outputs: $\mathbf{b} = [\mathbf{b}_s; \mathbf{b}_t] \in \mathbb{R}^{2p}$
\STATE \hspace{1em} Element-wise multiply with trunk output: $\mathbf{c} = \mathbf{b} \odot \mathbf{t}$ \hfill $\triangleright$ DeepONet operator\smallskip
\STATE \textbf{5. Shared Decoder:} Pass $\mathbf{c}$ through MLP decoder to hidden representation $\mathbf{h} \in \mathbb{R}^{d_h}$\smallskip
\STATE \textbf{6. Output Heads:} 
\STATE \hspace{1em} Spatial output: $\hat{\mathbf{u}}_s =$ linear head mapping $\mathbf{h} \rightarrow \mathbb{R}^{H \times W}$
\STATE \hspace{1em} Temporal output: $\hat{\mathbf{u}}_t =$ linear head mapping $\mathbf{h} \rightarrow \mathbb{R}^{T}$\smallskip
\STATE \textbf{return} $\hat{\mathbf{u}}_s, \hat{\mathbf{u}}_t$
\end{algorithmic}
\end{algorithm}

\section{Experimental Methods}\label{app:methods}
\subsection{SSFM}
SSFM is used to numerically integrate \eqref{eq:gnlse}. The method splits the propagation over a small step \(dz\) into two sub-steps: a linear step where dispersion and diffraction are applied in Fourier space, and a nonlinear step where the Kerr effect is applied in the time domain:

\begin{equation}
	A(z + dz) \approx e^{\hat{D} dz/2} \; e^{i \gamma |A(z)|^2 dz} \; e^{\hat{D} dz/2} \; A(z).
\end{equation}
This symmetric splitting ensures second-order accuracy in the step size \(dz\) and allows efficient computation using fast Fourier transforms (FFTs) for the linear operator. By iterating this procedure along \(z\), the evolution of the full spatio-temporal field can be obtained~\cite{agrawal2013nonlinear,teugin2021reusability}.

\subsection{Spectral convolution 2D}
Spectral convolution 2D applies convolution in the Fourier domain to a 2D function \(u(x,y)\). Its Fourier transform is
\begin{equation}
	\hat{u}(k_x, k_y) = \sum_{x=0}^{N_x-1} \sum_{y=0}^{N_y-1} u(x,y) \, e^{-2\pi i \left(\frac{k_x x}{N_x} + \frac{k_y y}{N_y}\right)},
\end{equation}
A subset of Fourier modes \((k_x, k_y) \in \mathcal{K}_x \times \mathcal{K}_y\) is selected, and learnable complex weights \(W(k_x, k_y)\) are applied:
\begin{equation}
	\hat{v}(k_x, k_y) = W(k_x, k_y) \cdot \hat{u}(k_x, k_y), \quad (k_x, k_y) \in \mathcal{K}_x \times \mathcal{K}_y.
\end{equation}
The output in the original domain is obtained via the inverse Fourier transform:
\begin{equation}
	v(x,y) = \sum_{(k_x, k_y) \in \mathcal{K}_x \times \mathcal{K}_y} \hat{v}(k_x, k_y) \, e^{2\pi i \left(\frac{k_x x}{N_x} + \frac{k_y y}{N_y}\right)}.
\end{equation}\cite{li2021fourierneuraloperatorparametric}

\subsection{Spectral convolution 1D}
Spectral convolution 1D applies convolution in the Fourier domain to a function \(u(t)\). Its Fourier transform is
\begin{equation}
	\hat{u}(k) = \sum_{t=0}^{L-1} u(t) \, e^{-2\pi i \frac{k t}{L}},
\end{equation}
A subset of Fourier modes \(k \in \mathcal{K}\) is selected, and learnable complex weights \(W(k)\) are applied:
\begin{equation}
	\hat{v}(k) = W(k) \cdot \hat{u}(k), \quad k \in \mathcal{K}.
\end{equation}
The output in the original domain is obtained via the inverse Fourier transform:
\begin{equation}
	v(t) = \sum_{k \in \mathcal{K}} \hat{v}(k) \, e^{2\pi i \frac{k t}{L}}.
\end{equation}\cite{li2021fourierneuraloperatorparametric}

\subsection{Fourier Embedding}
Fourier Embedding maps a low-dimensional input \(\mathbf{s} \in \mathbb{R}^{d}\) into a higher-dimensional space using a set of sinusoidal functions. This allows the network to represent high-frequency variations and oscillatory dependencies efficiently. The mapping is defined as
\begin{equation}
	\mathbf{\Phi}(\mathbf{s}) = \big[ \, \sin(2 \pi \mathbf{B} \mathbf{s}), \; \cos(2 \pi \mathbf{B} \mathbf{s}) \, \big],
\end{equation}
where \(\mathbf{B} \in \mathbb{R}^{(d \times D/2)}\) is a learnable or randomly initialized matrix and \(D\) is the embedding dimension. Each component of \(\mathbf{s}\) is projected through \(\mathbf{B}\) to produce a set of periodic features:
\begin{equation}
	\begin{split}
		\mathbf{s} \mapsto &[\, \sin(2 \pi \sum_j B_{jk} s_j), \; \cos(2 \pi \sum_j B_{jk} s_j) \,] \\
		&\quad \text{for } k=1,\dots,D/2.
	\end{split}
\end{equation}
This embedding enhances the expressiveness of the network, enabling it to capture nonlinear and high-frequency interactions in the input space~\cite{10.5555/3495724.3496356}.

\subsection{Linear layer}
Linear layer applies an affine transformation to an input vector \(\mathbf{x} \in \mathbb{R}^{d_{\text{in}}}\). The operation is defined as
\begin{equation}
	\mathbf{y} = \mathbf{W} \mathbf{x} + \mathbf{b},
\end{equation}
where \(\mathbf{W} \in \mathbb{R}^{d_{\text{out}} \times d_{\text{in}}}\) is a learnable weight matrix and 
\(\mathbf{b} \in \mathbb{R}^{d_{\text{out}}}\) is a learnable bias vector. Each output component is therefore a weighted sum of the input components:
\begin{equation}
	y_i = \sum_{j=1}^{d_{\text{in}}} W_{ij} x_j + b_i,
	\qquad i = 1,\dots,d_{\text{out}}.
\end{equation}
The linear layer provides a fundamental building block for neural networks, enabling arbitrary affine mappings between vector spaces~\cite{Goodfellow-et-al-2016,bishop2023deep}.

\subsection{Normalization layer}
Normalization layer mproves numerical stability by rescaling activations using shared statistics.  
Given a set of activations \(x_i\), the normalized output is
\begin{equation}
	\hat{x}_i = \frac{x_i - \mu}{\sigma},
\end{equation}
where \(\mu\) and \(\sigma\) denote the mean and standard deviation computed over a chosen set of elements.  
The choice of normalization domain determines the variant: the statistics may be computed across a batch of samples (batch normalization), or across the features of a single sample (layer normalization)~\cite{Goodfellow-et-al-2016,bishop2023deep}.

\subsection{Activation function}
Activation function introduces nonlinearity into neural networks, allowing them to approximate complex mappings beyond what is possible with purely linear operations.  
Given an input \(x\), an activation function applies a pointwise transformation
\begin{equation}
	y = \phi(x),
\end{equation}
which shapes how information flows through the network and determines its expressive power.  
\noindent
In this work, we use the \textbf{Re}ctified \textbf{L}inear \textbf{U}nit (ReLU), defined by
\begin{equation}
	\phi(x) = \max(0, x),
\end{equation}
a widely adopted activation due to its simplicity, numerical stability, and effectiveness in deep architectures~\cite{Goodfellow-et-al-2016,bishop2023deep}.

\subsection{Pooling}
Pooling is an operation that reduces the spatial resolution of feature maps by summarizing information within small local windows. It helps retain the most important structural patterns while decreasing dimensionality, making subsequent processing more efficient. In this work, max pooling was used, where each window is replaced by its maximum value~\cite{Goodfellow-et-al-2016,bishop2023deep,10.5555/248702}.

\subsection{AdamW}
\begin{algorithm}[!ht]
\caption{AdamW Algorithm \cite{kingma2017adammethodstochasticoptimization,loshchilov2019decoupled}}
\begin{algorithmic}[1]
\STATE \textbf{given} $\alpha = 5 \times 10^{-4}$, $\beta_1 = 0.9$, $\beta_2 = 0.999$, $\epsilon = 10^{-8}$, $\lambda \in \mathbb{R}$
\STATE \textbf{initialize} time step $t \leftarrow 0$, parameter vector $\theta_{t=0} \in \mathbb{R}^n$, first moment vector $m_{t=0} \leftarrow \theta$, second moment vector $v_{t=0} \leftarrow \theta$, schedule multiplier $\eta_{t=0} \in \mathbb{R}$
\REPEAT
\STATE $t \leftarrow t + 1$
\STATE $\nabla f_t(\theta_{t-1}) \leftarrow \text{SelectBatch}(\theta_{t-1})$ \hfill $\triangleright$ select batch and return the corresponding gradient
\STATE $g_t \leftarrow \nabla f_t(\theta_{t-1})$
\STATE $m_t \leftarrow \beta_1 m_{t-1} + (1 - \beta_1)g_t$ \hfill $\triangleright$ here and below all operations are element-wise
\STATE $v_t \leftarrow \beta_2 v_{t-1} + (1 - \beta_2)g_t^2$
\STATE $\hat{m}_t \leftarrow m_t/(1 - \beta_1^t)$ \hfill $\triangleright$ $\beta_1$ is taken to the power of $t$
\STATE $\hat{v}_t \leftarrow v_t/(1 - \beta_2^t)$ \hfill $\triangleright$ $\beta_2$ is taken to the power of $t$
\STATE $\eta_t \leftarrow \text{SetScheduleMultiplier}(t)$ \hfill $\triangleright$ can be fixed, decay, or also be used for warm restarts
\STATE $\theta_t \leftarrow \theta_{t-1} - \eta_t\left(\alpha\hat{m}_t/(\sqrt{\hat{v}_t} + \epsilon) + \lambda\theta_{t-1}\right)$
\UNTIL{stopping criterion is met}
\STATE \textbf{return} optimized parameters $\theta_t$
\end{algorithmic}
\end{algorithm}

\subsection{Gradient clipping}
Gradient clipping is a technique used during training to bound the size of gradients and prevent unstable or exploding updates. When using an $L_2$ norm threshold, the gradient vector $\mathbf{g}$ is rescaled if its norm exceeds a maximum value (here, 1.0). Formally, if $\|\mathbf{g}\|_2 > 1.0$, then
\begin{equation}
    \mathbf{g} \leftarrow \frac{\mathbf{g}}{\|\mathbf{g}\|_2}\times 1.0,
\end{equation}
ensuring that the $L_2$ norm of the gradient does not exceed 1.0 while preserving its direction. This stabilizes training by preventing excessively large parameter updates, which is especially useful during backpropagation in deep learning optimization~\cite{Goodfellow-et-al-2016,bishop2023deep, zhang2020gradientclipping}.

\section*{Acknowledgment}
The authors acknowledge

\section*{Conflict of Interest}
The authors declare no conflict of interest.

\bibliographystyle{unsrt}
\bibliography{references} 

@book{agrawal2013nonlinear,
	title={Nonlinear Fiber Optics},
	author={Agrawal, G.P.},
	isbn={9780123970237},
	lccn={2013431001},
	series={Optics and Photonics},
	url={https://books.google.co.in/books?id=xNvw-GDVn84C},
	year={2013},
	publisher={Elsevier Science}
}

@article{AGRAWAL2019309,
	title = {Invite paper: Self-imaging in multimode graded-index fibers and its impact on the nonlinear phenomena},
	journal = {Optical Fiber Technology},
	volume = {50},
	pages = {309-316},
	year = {2019},
	issn = {1068-5200},
	doi = {https://doi.org/10.1016/j.yofte.2019.04.012},
	url = {https://www.sciencedirect.com/science/article/pii/S106852001930104X},
	author = {Govind P. Agrawal}
}

@article{Wright2015,
	author = {Wright, Logan G. and Christodoulides, Demetrios N. and Wise, Frank W.},
	title = {Controllable spatiotemporal nonlinear effects in multimode fibres},
	journal = {Nature Photonics},
	volume = {9},
	number = {5},
	pages = {306--310},
	year = {2015},
	month = {May},
	doi = {10.1038/nphoton.2015.61},
	issn = {1749-4893},
	url = {https://doi.org/10.1038/nphoton.2015.61}
}

@article{krupa2016observation,
	title={Observation of geometric parametric instability induced by the periodic spatial self-imaging of multimode waves},
	author={Krupa, Katarzyna and Tonello, Alessandro and Barth{\'e}l{\'e}my, Alain and Couderc, Vincent and Shalaby, Badr Mohamed and Bendahmane, Abdelkrim and Millot, Guy and Wabnitz, Stefan},
	journal={Physical review letters},
	volume={116},
	number={18},
	pages={183901},
	year={2016},
	publisher={APS}
}

@ARTICLE{8094262,
	author={Teğin, Uğur and Ortaç, Bülend},
	journal={IEEE Photonics Technology Letters}, 
	title={Spatiotemporal Instability of Femtosecond Pulses in Graded-Index Multimode Fibers}, 
	year={2017},
	volume={29},
	number={24},
	pages={2195-2198},
	doi={10.1109/LPT.2017.2769343}
}

@incollection{krupa2017spatiotemporal,
	title={Spatiotemporal nonlinear dynamics in multimode fibers},
	author={Krupa, K and Couderc, V and Tonello, A and Picozzi, A and Barth{\'e}l{\'e}my, A and Millot, G and Wabnitz, S},
	booktitle={Nonlinear Guided Wave Optics: A testbed for extreme waves},
	pages={14--1},
	year={2017},
	publisher={IOP Publishing Bristol, UK}
}

@article{10.1063/1.5119434,
	author = {Krupa, Katarzyna and Tonello, Alessandro and Barthélémy, Alain and Mansuryan, Tigran and Couderc, Vincent and Millot, Guy and Grelu, Philippe and Modotto, Daniele and Babin, Sergey A. and Wabnitz, Stefan},
	title = {Multimode nonlinear fiber optics, a spatiotemporal avenue},
	journal = {APL Photonics},
	volume = {4},
	number = {11},
	pages = {110901},
	year = {2019},
	month = {11},
	issn = {2378-0967},
	doi = {10.1063/1.5119434},
	url = {https://doi.org/10.1063/1.5119434},
	eprint = {https://pubs.aip.org/aip/app/article-pdf/doi/10.1063/1.5119434/19994365/110901_1_1.5119434.pdf},
}

@article{Hansson:20,
	author = {Tobias Hansson and Alessandro Tonello and Tigran Mansuryan and Fabio Mangini and Mario Zitelli and Mario Ferraro and Alioune Niang and Rocco Crescenzi and Stefan Wabnitz and Vincent Couderc},
	journal = {Opt. Express},
	number = {16},
	pages = {24005--24021},
	publisher = {Optica Publishing Group},
	title = {Nonlinear beam self-imaging and self-focusing dynamics in a GRIN multimode optical fiber: theory and experiments},
	volume = {28},
	month = {Aug},
	year = {2020},
	url = {https://opg.optica.org/oe/abstract.cfm?URI=oe-28-16-24005},
	doi = {10.1364/OE.398531},
}

@article{Krupa_2017,
	title={Spatial beam self-cleaning in multimode fibres},
	volume={11},
	ISSN={1749-4893},
	url={http://dx.doi.org/10.1038/nphoton.2017.32},
	DOI={10.1038/nphoton.2017.32},
	number={4},
	journal={Nature Photonics},
	publisher={Springer Science and Business Media LLC},
	author={Krupa, K. and Tonello, A. and Shalaby, B. M. and Fabert, M. and Barthélémy, A. and Millot, G. and Wabnitz, S. and Couderc, V.},
	year={2017},
	month=mar, pages={237–241} }

@article{Ai:17,
	author = {Min Ai and Weihang Shu and Tim Salcudean and Robert Rohling and Purang Abolmaesumi and Shuo Tang},
	journal = {Opt. Express},
	number = {15},
	pages = {17713--17726},
	publisher = {Optica Publishing Group},
	title = {Design of high energy laser pulse delivery in a multimode fiber for photoacoustic tomography},
	volume = {25},
	month = {Jul},
	year = {2017},
	url = {https://opg.optica.org/oe/abstract.cfm?URI=oe-25-15-17713},
	doi = {10.1364/OE.25.017713},
}

@article{Morales-Delgado:15,
	author = {Edgar E. Morales-Delgado and Demetri Psaltis and Christophe Moser},
	journal = {Opt. Express},
	number = {25},
	pages = {32158--32170},
	publisher = {Optica Publishing Group},
	title = {Two-photon imaging through a multimode fiber},
	volume = {23},
	month = {Dec},
	year = {2015},
	url = {https://opg.optica.org/oe/abstract.cfm?URI=oe-23-25-32158},
	doi = {10.1364/OE.23.032158},
}

@article{Gusachenko2017,
	author = {Gusachenko, Ivan and Nylk, Jonathan and Tello, Javier A and Dholakia, Kishan},
	title = {Multimode fibre based imaging for optically cleared samples},
	journal = {Biomedical Optics Express},
	volume = {8},
	number = {11},
	pages = {5179--5190},
	year = {2017},
	month = {Nov},
	day = {1},
	publisher = {Optical Society of America},
	doi = {10.1364/BOE.8.005179},
	pmid = {29188112},
	pmcid = {PMC5695962},
	url = {https://doi.org/10.1364/BOE.8.005179}
}

@article{
	doi:10.1126/science.aao0831,
	author = {Logan G. Wright  and Demetrios N. Christodoulides  and Frank W. Wise },
	title = {Spatiotemporal mode-locking in multimode fiber lasers},
	journal = {Science},
	volume = {358},
	number = {6359},
	pages = {94-97},
	year = {2017},
	doi = {10.1126/science.aao0831},
	URL = {https://www.science.org/doi/abs/10.1126/science.aao0831},
	eprint = {https://www.science.org/doi/pdf/10.1126/science.aao0831}
}

@article{PhysRevLett.126.093901,
	title = {Spatiotemporal Mode-Locking in Lasers with Large Modal Dispersion},
	author = {Ding, Yihang and Xiao, Xiaosheng and Liu, Kewei and Fan, Shuzheng and Zhang, Xiaoguang and Yang, Changxi},
	journal = {Phys. Rev. Lett.},
	volume = {126},
	issue = {9},
	pages = {093901},
	numpages = {6},
	year = {2021},
	month = {Mar},
	publisher = {American Physical Society},
	doi = {10.1103/PhysRevLett.126.093901},
	url = {https://link.aps.org/doi/10.1103/PhysRevLett.126.093901}
}

@article{Cao2023,
	author = {Cao, Bo and Gao, Chenxin and Liu, Kewei and Xiao, Xiaosheng and Yang, Changxi and Bao, Chengying},
	title = {Spatiotemporal mode-locking and dissipative solitons in multimode fiber lasers},
	journal = {Light: Science \& Applications},
	volume = {12},
	number = {1},
	pages = {260},
	year = {2023},
	month = {Oct},
	day = {30},
	doi = {10.1038/s41377-023-01305-0},
	issn = {2047-7538},
	url = {https://doi.org/10.1038/s41377-023-01305-0}
}

@article{Paudel_2020,
	title={Classification of time-domain waveforms using a speckle-based optical reservoir computer},
	volume={28},
	ISSN={1094-4087},
	url={http://dx.doi.org/10.1364/OE.379264},
	DOI={10.1364/oe.379264},
	number={2},
	journal={Optics Express},
	publisher={Optica Publishing Group},
	author={Paudel, Uttam and Luengo-Kovac, Marta and Pilawa, Jacob and Shaw, T. Justin and Valley, George C.},
	year={2020},
	month=jan, pages={1225} 
}

@inproceedings{Rahmani:21,
	author = {Babak Rahmani and Ugur Tegin and Mustafa Y{i}ld{i}r{i}m and \.{I}lker O\u{g}uz and Damien Loterie and Eirini Kakkava and Navid Borhani and Demetri Psaltis and Christophe Moser},
	booktitle = {Optical Fiber Communication Conference (OFC) 2021},
	journal = {Optical Fiber Communication Conference (OFC) 2021},
	pages = {Th5B.1},
	publisher = {Optica Publishing Group},
	title = {Learning to See and Compute through Multimode Fibers},
	year = {2021},
	url = {https://opg.optica.org/abstract.cfm?URI=OFC-2021-Th5B.1},
	doi = {10.1364/OFC.2021.Th5B.1},
}

@book{kolesik2017computational,
	title={Computational Optics: Beam Propagation Methods},
	author={Kolesik, M.},
	isbn={9781536123340},
	lccn={2017947971},
	series={Physics research and technology},
	url={https://books.google.co.in/books?id=GeOzswEACAAJ},
	year={2017},
	publisher={Nova Science Publishers}
}

@article{Brehler_2020,
	title={Simulation of nonlinear signal propagation in multimode fibers on multi-GPU systems},
	volume={84},
	ISSN={1007-5704},
	url={http://dx.doi.org/10.1016/j.cnsns.2019.105150},
	DOI={10.1016/j.cnsns.2019.105150},
	journal={Communications in Nonlinear Science and Numerical Simulation},
	publisher={Elsevier BV},
	author={Brehler, Marius and Schirwon, Malte and Krummrich, Peter M. and Göddeke, Dominik},
	year={2020},
	month=may, pages={105150} }

@INPROCEEDINGS{8437734,
	author={Häger, Christian and Pfister, Henry D.},
	booktitle={2018 IEEE International Symposium on Information Theory (ISIT)}, 
	title={Deep Learning of the Nonlinear Schrödinger Equation in Fiber-Optic Communications}, 
	year={2018},
	volume={},
	number={},
	pages={1590-1594},
	doi={10.1109/ISIT.2018.8437734}
}

@article{Genty2021,
	author = {Goëry Genty and Lauri Salmela and John M. Dudley and Daniel Brunner and Alexey Kokhanovskiy and Sergei Kobtsev and Sergei K. Turitsyn},
	title = {Machine learning and applications in ultrafast photonics},
	journal = {Nature Photonics},
	volume = {15},
	number = {2},
	pages = {91-101},
	year = {2021},
	doi = {10.1038/s41566-020-00716-4},
	url = {https://doi.org/10.1038/s41566-020-00716-4}
}

@INPROCEEDINGS{9596142,
	author={Stucchi, Diego and Corsini, Andrea and Genty, Goëry and Boracchi, Giacomo and Foi, Alessandro},
	booktitle={2021 IEEE 31st International Workshop on Machine Learning for Signal Processing (MLSP)}, 
	title={A Weighted Loss Function to Predict Control Parameters for Supercontinuum Generation Via Neural Networks}, 
	year={2021},
	volume={},
	number={},
	pages={1-6},
	doi={10.1109/MLSP52302.2021.9596142}
}

@article{Salmela2021,
	author    = {Salmela, Lauri and Tsipinakis, Nikolaos and Foi, Alessandro and Billet, Cyril and Dudley, John M. and Genty, Goëry},
	title     = {Predicting ultrafast nonlinear dynamics in fibre optics with a recurrent neural network},
	journal   = {Nature Machine Intelligence},
	year      = {2021},
	volume    = {3},
	number    = {4},
	pages     = {344--354},
	doi       = {10.1038/s42256-021-00297-z},
	url       = {https://doi.org/10.1038/s42256-021-00297-z}
}

@article{teugin2021reusability,
    title={Reusability report: Predicting spatiotemporal nonlinear dynamics in multimode fibre optics with a recurrent neural network},
    author={Te{\u{g}}in, U{\u{g}}ur and Din{\c{c}}, Niyazi Ula{\c{s}} and Moser, Christophe and Psaltis, Demetri},
    journal={Nature Machine Intelligence},
    pages={1--5},
    year={2021},
    publisher={Nature Publishing Group}
}

@article{Chen2023,
    author = {Chen, Zaijun and Sludds, Alexander and Davis, Ronald and Christen, Ian and Bernstein, Liane and Ateshian, Lamia and Heuser, Tobias and Heermeier, Niels and Lott, James A. and Reitzenstein, Stephan and Hamerly, Ryan and Englund, Dirk},
    title = {Deep learning with coherent VCSEL neural networks},
    journal = {Nature Photonics},
    volume = {17},
    number = {8},
    pages = {723--730},
    year = {2023},
    month = {Aug},
    doi = {10.1038/s41566-023-01233-w},
    issn = {1749-4893},
    url = {https://doi.org/10.1038/s41566-023-01233-w}
}

@article{Fang:23,
    author = {Yin Fang and Hao-Bin Han and Wen-Bo Bo and Wei Liu and Ben-Hai Wang and Yue-Yue Wang and Chao-Qing Dai},
    journal = {Opt. Lett.},
    number = {3},
    pages = {779--782},
    publisher = {Optica Publishing Group},
    title = {Deep neural network for modeling soliton dynamics in the mode-locked laser},
    volume = {48},
    month = {Feb},
    year = {2023},
    url = {https://opg.optica.org/ol/abstract.cfm?URI=ol-48-3-779},
    doi = {10.1364/OL.482946},
}

@article{Hadad_2023,
    doi = {10.1088/2040-8986/ad08dc},
    url = {https://doi.org/10.1088/2040-8986/ad08dc},
    year = {2023},
    month = {nov},
    publisher = {IOP Publishing},
    volume = {25},
    number = {12},
    pages = {123501},
    author = {Hadad, Barak and Froim, Sahar and Yosef, Erez and Giryes, Raja and Bahabad, Alon},
    title = {Deep learning in optics—a tutorial},
    journal = {Journal of Optics}
}

@article{10.1063/5.0169810,
    author = {Tsakyridis, Apostolos and Moralis-Pegios, Miltiadis and Giamougiannis, George and Kirtas, Manos and Passalis, Nikolaos and Tefas, Anastasios and Pleros, Nikos},
    title = {Photonic neural networks and optics-informed deep learning fundamentals},
    journal = {APL Photonics},
    volume = {9},
    number = {1},
    pages = {011102},
    year = {2024},
    month = {01},
    issn = {2378-0967},
    doi = {10.1063/5.0169810},
    url = {https://doi.org/10.1063/5.0169810},
    eprint = {https://pubs.aip.org/aip/app/article-pdf/doi/10.1063/5.0169810/19848606/011102_1_5.0169810.pdf},
}

@misc{belonovskii2024vcsel,
    title={Predicting VCSEL Emission Properties Using Transformer Neural Networks}, 
    author={Aleksei V. Belonovskii and Elizaveta I. Girshova and Erkki Lähderanta and Mikhail Kaliteevski},
    year={2024},
    eprint={2407.06039},
    archivePrefix={arXiv},
    primaryClass={cond-mat.dis-nn},
    url={https://arxiv.org/abs/2407.06039}, 
}

@article{10.1002/adpr.202500149,
    author = {Murugan, Dinesh Kumar and Joseph, Rajkumar and Kanagaraj, Nithyanandan},
    title = {Transformer Encoder–Decoder Framework for Nonlinear Pulse Evolution and Inverse Modeling},
    journal = {Advanced Photonics Research},
    volume = {6},
    number = {11},
    pages = {2500149},
    doi = {https://doi.org/10.1002/adpr.202500149},
    url = {https://advanced.onlinelibrary.wiley.com/doi/abs/10.1002/adpr.202500149},
    eprint = {https://advanced.onlinelibrary.wiley.com/doi/pdf/10.1002/adpr.202500149},
    year = {2025}
}

@article{Freire:23,
	author = {Pedro Freire and Egor Manuylovich and Jaroslaw E. Prilepsky and Sergei K. Turitsyn},
	journal = {Adv. Opt. Photon.},
	number = {3},
	pages = {739--834},
	publisher = {Optica Publishing Group},
	title = {Artificial neural networks for photonic applications---from algorithms to implementation: tutorial},
	volume = {15},
	month = {Sep},
	year = {2023},
	url = {https://opg.optica.org/aop/abstract.cfm?URI=aop-15-3-739},
	doi = {10.1364/AOP.484119},
}

@article{RAISSI2019686,
    title = {Physics-informed neural networks: A deep learning framework for solving forward and inverse problems involving nonlinear partial differential equations},
    journal = {Journal of Computational Physics},
    volume = {378},
    pages = {686-707},
    year = {2019},
    issn = {0021-9991},
    doi = {https://doi.org/10.1016/j.jcp.2018.10.045},
    url = {https://www.sciencedirect.com/science/article/pii/S0021999118307125},
    author = {M. Raissi and P. Perdikaris and G.E. Karniadakis}
}

@article{Lu2021,
    author = {Lu, Lu and Jin, Pengzhan and Pang, Guofei and Zhang, Zhongqiang and Karniadakis, George Em},
    title = {Learning nonlinear operators via DeepONet based on the universal approximation theorem of operators},
    journal = {Nature Machine Intelligence},
    volume = {3},
    number = {3},
    pages = {218--229},
    year = {2021},
    month = {Mar},
    doi = {10.1038/s42256-021-00302-5},
    issn = {2522-5839},
    url = {https://doi.org/10.1038/s42256-021-00302-5}
}

@misc{li2021fourierneuraloperatorparametric,
    title={Fourier Neural Operator for Parametric Partial Differential Equations}, 
    author={Zongyi Li and Nikola Kovachki and Kamyar Azizzadenesheli and Burigede Liu and Kaushik Bhattacharya and Andrew Stuart and Anima Anandkumar},
    year={2021},
    eprint={2010.08895},
    archivePrefix={arXiv},
    primaryClass={cs.LG},
    url={https://arxiv.org/abs/2010.08895}, 
}

@inproceedings{10.5555/3495724.3496356,
    author = {Tancik, Matthew and Srinivasan, Pratul P. and Mildenhall, Ben and Fridovich-Keil, Sara and Raghavan, Nithin and Singhal, Utkarsh and Ramamoorthi, Ravi and Barron, Jonathan T. and Ng, Ren},
    title = {Fourier features let networks learn high frequency functions in low dimensional domains},
    year = {2020},
    isbn = {9781713829546},
    publisher = {Curran Associates Inc.},
    address = {Red Hook, NY, USA},
    booktitle = {Proceedings of the 34th International Conference on Neural Information Processing Systems},
    articleno = {632},
    numpages = {11},
    location = {Vancouver, BC, Canada},
    series = {NIPS '20}
}

@ARTICLE{9815178,
    author={Wang, Danshi and Jiang, Xiaotian and Song, Yuchen and Fu, Meixia and Zhang, Zhiguo and Chen, Xue and Zhang, Min},
    journal={IEEE Communications Magazine}, 
    title={Applications of Physics-Informed Neural Network for Optical Fiber Communications}, 
    year={2022},
    volume={60},
    number={9},
    pages={32-37},
    doi={10.1109/MCOM.001.2100961}
}

@article{MARGENBERG2024112725,
    title = {Optimal Dirichlet boundary control by Fourier neural operators applied to nonlinear optics},
    journal = {Journal of Computational Physics},
    volume = {499},
    pages = {112725},
    year = {2024},
    issn = {0021-9991},
    doi = {https://doi.org/10.1016/j.jcp.2023.112725},
    url = {https://www.sciencedirect.com/science/article/pii/S0021999123008203},
    author = {Nils Margenberg and Franz X. Kärtner and Markus Bause}
}

@misc{long2025invertiblefourierneuraloperators,
    title={Invertible Fourier Neural Operators for Tackling Both Forward and Inverse Problems}, 
    author={Da Long and Zhitong Xu and Qiwei Yuan and Yin Yang and Shandian Zhe},
    year={2025},
    eprint={2402.11722},
    archivePrefix={arXiv},
    primaryClass={cs.LG},
    url={https://arxiv.org/abs/2402.11722}, 
}

@ARTICLE{10195160,
    author={Zhang, Ximeng and Zhang, Min and Song, Yuchen and Jiang, Xiaotian and Zhang, Fan and Wang, Danshi},
    journal={Journal of Lightwave Technology}, 
    title={DeepONet-Based Waveform-Level Simulation for a Wideband Nonlinear WDM System}, 
    year={2023},
    volume={41},
    number={22},
    pages={6908-6922},
    doi={10.1109/JLT.2023.3298881}
}

@article{10.1364/AOP.484298,
    author = {Hui Cao and Tom\'{a}\v{s} \v{C}i\v{z}m\'{a}r and Sergey Turtaev and Tom\'{a}\v{s} Tyc and Stefan Rotter},
    journal = {Adv. Opt. Photon.},
    number = {2},
    pages = {524--612},
    publisher = {Optica Publishing Group},
    title = {Controlling light propagation in multimode fibers for imaging, spectroscopy, and beyond},
    volume = {15},
    month = {Jun},
    year = {2023},
    url = {https://opg.optica.org/aop/abstract.cfm?URI=aop-15-2-524},
    doi = {10.1364/AOP.484298},
}

@article{10.1117/1.APN.2.6.066005,
    author = {Shengfu Cheng and Xuyu Zhang and Tianting Zhong and Huanhao Li and Haoran Li and Lei Gong and Honglin Liu and Puxiang Lai},
    title = {{Nonconvex optimization for optimum retrieval of the transmission matrix of a multimode fiber}},
    volume = {2},
    journal = {Advanced Photonics Nexus},
    number = {6},
    publisher = {SPIE},
    pages = {066005},
    year = {2023},
    doi = {10.1117/1.APN.2.6.066005},
    URL = {https://doi.org/10.1117/1.APN.2.6.066005}
}

@article{Mafi:10,
    author = {Arash Mafi},
    journal = {J. Lightwave Technol.},
    number = {10},
    pages = {1547--1555},
    publisher = {Optica Publishing Group},
    title = {Bandwidth Improvement in Multimode Optical Fibers Via Scattering From Core Inclusions},
    volume = {28},
    month = {Mar},
    year = {2010},
    url = {https://opg.optica.org/jlt/abstract.cfm?URI=jlt-28-10-1547},
}

@book{snyder1983optical,
    title={Optical Waveguide Theory},
    author={Snyder, A.W. and Love, J.},
    isbn={9780412099502},
    lccn={lc83007463},
    series={Science paperbacks},
    url={https://books.google.co.in/books?id=gIQB_hzB0SMC},
    year={1983},
    publisher={Springer US}
}

@book{agrawal2010applications,
    title={Applications of Nonlinear Fiber Optics},
    author={Agrawal, G.P.},
    isbn={9780080568768},
    url={https://books.google.co.in/books?id=oy_CDAAAQBAJ},
    year={2010},
    publisher={Academic Press}
}

@article{MAHARANA202291,
    title = {A review: Data pre-processing and data augmentation techniques},
    journal = {Global Transitions Proceedings},
    volume = {3},
    number = {1},
    pages = {91-99},
    year = {2022},
    note = {International Conference on Intelligent Engineering Approach(ICIEA-2022)},
    issn = {2666-285X},
    doi = {https://doi.org/10.1016/j.gltp.2022.04.020},
    url = {https://www.sciencedirect.com/science/article/pii/S2666285X22000565},
    author = {Kiran Maharana and Surajit Mondal and Bhushankumar Nemade}
}

@book{bishop2023deep,
    title={Deep Learning: Foundations and Concepts},
    author={Bishop, C.M. and Bishop, H.},
    isbn={9783031454684},
    series={Computer Science},
    url={https://books.google.co.in/books?id=0uTgEAAAQBAJ},
    year={2023},
    publisher={Springer International Publishing}
}

@book{Goodfellow-et-al-2016,
    title={Deep Learning},
    author={Ian Goodfellow and Yoshua Bengio and Aaron Courville},
    publisher={MIT Press},
    note={\url{http://www.deeplearningbook.org}},
    year={2016}
}

@book{10.5555/248702,
    author = {Oppenheim, Alan V. and Willsky, Alan S. and Nawab, S. Hamid},
    title = {Signals \& systems (2nd ed.)},
    year = {1996},
    isbn = {0138147574},
    publisher = {Prentice-Hall, Inc.},
    address = {USA}
}

@misc{kingma2017adammethodstochasticoptimization,
    title={Adam: A Method for Stochastic Optimization}, 
    author={Diederik P. Kingma and Jimmy Ba},
    year={2017},
    eprint={1412.6980},
    archivePrefix={arXiv},
    primaryClass={cs.LG},
    url={https://arxiv.org/abs/1412.6980}, 
}

@misc{loshchilov2019decoupled,
    title={Decoupled Weight Decay Regularization}, 
    author={Ilya Loshchilov and Frank Hutter},
    year={2019},
    eprint={1711.05101},
    archivePrefix={arXiv},
    primaryClass={cs.LG},
    url={https://arxiv.org/abs/1711.05101}, 
}

@misc{zhang2020gradientclipping,
      title={Why gradient clipping accelerates training: A theoretical justification for adaptivity}, 
      author={Jingzhao Zhang and Tianxing He and Suvrit Sra and Ali Jadbabaie},
      year={2020},
      eprint={1905.11881},
      archivePrefix={arXiv},
      primaryClass={math.OC},
      url={https://arxiv.org/abs/1905.11881}, 
}


%\vfill


\end{document}